\documentclass[preprint,12pt, 5p]{elsarticle}

\usepackage{graphicx}
\usepackage{dcolumn}
\usepackage{bm}
\usepackage{amsmath}
\usepackage{amsfonts}

\usepackage{xcolor}
\usepackage{paralist}
\usepackage{colortbl}
\usepackage{siunitx}
\usepackage{hyperref}
\usepackage{subfig}
\usepackage{listings}

\usepackage{booktabs}

\journal{Nuclear Instruments and Methods in Physics Research Section B}

\begin{document}
\begin{frontmatter}



\title{Neutron Yield Predictions with Artificial Neural Networks: \\ A Predictive Modeling Approach}


\author[inst1]{Benedikt Schmitz}

\affiliation[inst1]{organization={Technische Universität Darmstadt \\ Institut für Teilchenbeschleunigung und Elektromagnetische Felder (TEMF)},
            addressline={Schlossgartenstr.~8}, 
            city={Darmstadt},
            postcode={64289}, 
            country={Germany}}
\author[inst3]{Stefan Scheuren}
\affiliation[inst3]{organization={Technische Universität Darmstadt \\ Institut für Kernphysik (IKP)},
            addressline={Schlossgartenstr.~9}, 
            city={Darmstadt},
            postcode={64289}, 
            country={Germany}}
            
\begin{abstract}
The development of compact neutron sources for applications is extensive and features many approaches. 
Let alone ion-based approaches, several projects with different parameters exist.
This article focuses on ion-based neutron production below the spallation barrier for arbitrary light ion beams.
With this model, it is possible to compare different ion-based neutron source concepts against each other quickly.
This contribution derives a predictive model using Monte Carlo simulations (50k simulations) and deep neural networks. 
This model can skip the necessary Monte Carlo simulations, which individually take a long time to complete, increasing the effort for optimization and predictions.
The models' shortcomings are addressed, and mitigation strategies are proposed.  
\end{abstract}

\begin{highlights}
\item Fast inference model to predict neutron yields.
\item Model reproduces experimental data.
\item Model can compare between arbitrary ion sources.
\item Yield predictions can be adjusted and improved with each new data set.
\item Model and data are openly available.
\end{highlights}


\end{frontmatter}

\section{Introduction}
    The European Spallation Source ESS, the world's brightest neutron source, is nearing completion and further concentrates research with neutrons to only a few powerful facilities.
    This concentration is increased by the successive shutdown of European research reactors and counteracted by the development of compact sources at several institutions \cite{OTAKE2018,Ruecker2016,refId0}.
    \begin{figure}
        \centering
        \includegraphics{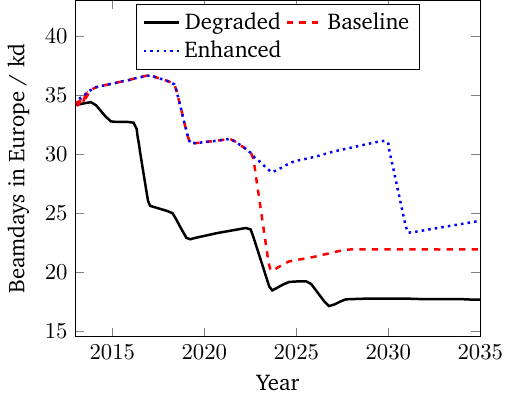}
        \caption{Prediction of the available beam days for neutron radiation in Europe. Three potential scenarios are displayed.
     Its differences are caused by the assumption of different remaining runtimes from existing machines and different commissions of new instruments, especially at the ESS.
     Enhanced includes faster commissioning of the ESS instruments and longer runtime, while Degraded assumes a faster decommissioning and delay in the Ess commissioning.
     The detailed explanation for each scenario is given in the ESFRI's report \cite[pp. 66-77]{ESFRI2016}.
     }
        \label{fig:neutronPredi}
    \end{figure}
    The loss of these sources further causes a decline in the available beam time as displayed in \autoref{fig:neutronPredi}.
    Compact neutron sources can also counteract this since several can be commissioned and used at the institutes where they are needed.

    \subsection{Physical constraints}
        These different concepts under development have in common that they use light ion projectiles (protons or deuterons) on a low mass target (e.g., lithium, beryllium\cite{OTAKE2018}, tantalum\cite{Ruecker2016}) and have particle energies below the spallation threshold($<\SI{100}{\mega\electronvolt}$( \cite[pp. 1008 f]{Conrad2021}. 
        Accelerator-based neutron sources are scalable, and for low-energy particles below the spallation threshold, nuclear waste is less of an issue, leaving compactness as the remaining issue.
        Compact accelerators generally mean that only lower energies can be achieved, which causes a reduction in the neutron yield.
        This makes a trade-off necessary for conventional accelerators since the accelerator can not become genuinely compact if high energies should be achieved, which increases the final neutron yield. 
        The different approaches, also mentioned in this article, are based on projectile energies lower than \SI{100}{\mega\electronvolt}\cite{Roth2016,international2021iaea,Ruecker2016,OTAKE2018,refId0}.
        Furthermore, higher neutron energies allow us to penetrate thicker materials and increase the number of applications for this source type, for example, in homeland security applications \cite{Buffler2004}.
        
        Models that ensure comparability along these different concepts and are quick to evaluate are essential for developing, comparing, and improving compact neutron sources.
        Models and results created by the individual institutions for their systems are not necessarily comparable since different nuclear cross-section libraries and measurement methods are used.

        Furthermore, our approach for a compact neutron source uses a laser-driven ion acceleration via the Target Normal Sheath Acceleration mechanism (TNSA)\cite{Roth2016}.
        TNSA ejects an exponential energy spectrum, which starkly contrasts the mono-energetic particles created by conventional accelerators.
    
    \subsection{Outline}
    In this work, we will present a model capable of evaluating a neutron source setup quickly and precisely in the relevant energy area and for arbitrary input bunches.

    Data-driven modeling needs a large dataset of consistent experimental data distributed over the parameter range of interest. 
    Due to the nonexistence of such a data set in the relevant energy range, simulations were used as the basis for modeling.
    The few data sets measured with similar parameters are used as test cases later in this work.
    
    The simulation results are used to devise a surrogate model based on artificial neural networks to predict the double differential neutron yield dynamically.
    For the uncertainty prediction, bootstrap training for the model was done. 
    
\section{Methods}
    \subsection{Simulations}
	We simulated the neutron yield with Monte Carlo Simulations utilizing the PHITS \cite{Sato2018} code in Version 3.28A with the FENDL \cite{FENDL3} library in Version 3. 

\begin{figure}
        \centering
        \subfloat[3D view of the simulation setup used in the PHITS simulations.\label{fig:MC-3d-setup}]{\includegraphics[width=0.48\textwidth]{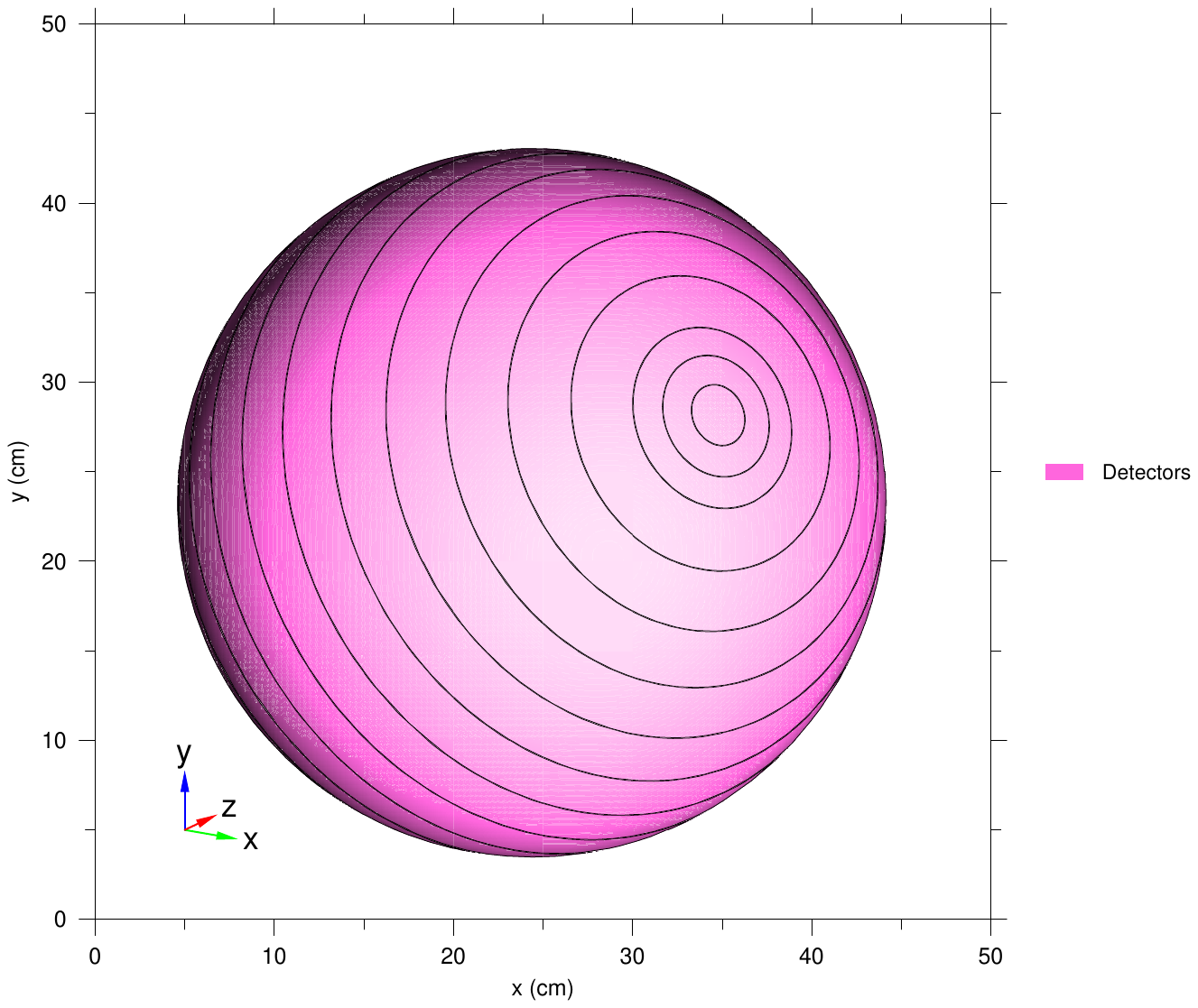}}\\%
        \subfloat[2D Cross section of the simulation geometry. \label{fig:MC-crossSection}]{\includegraphics[width=0.48\textwidth]{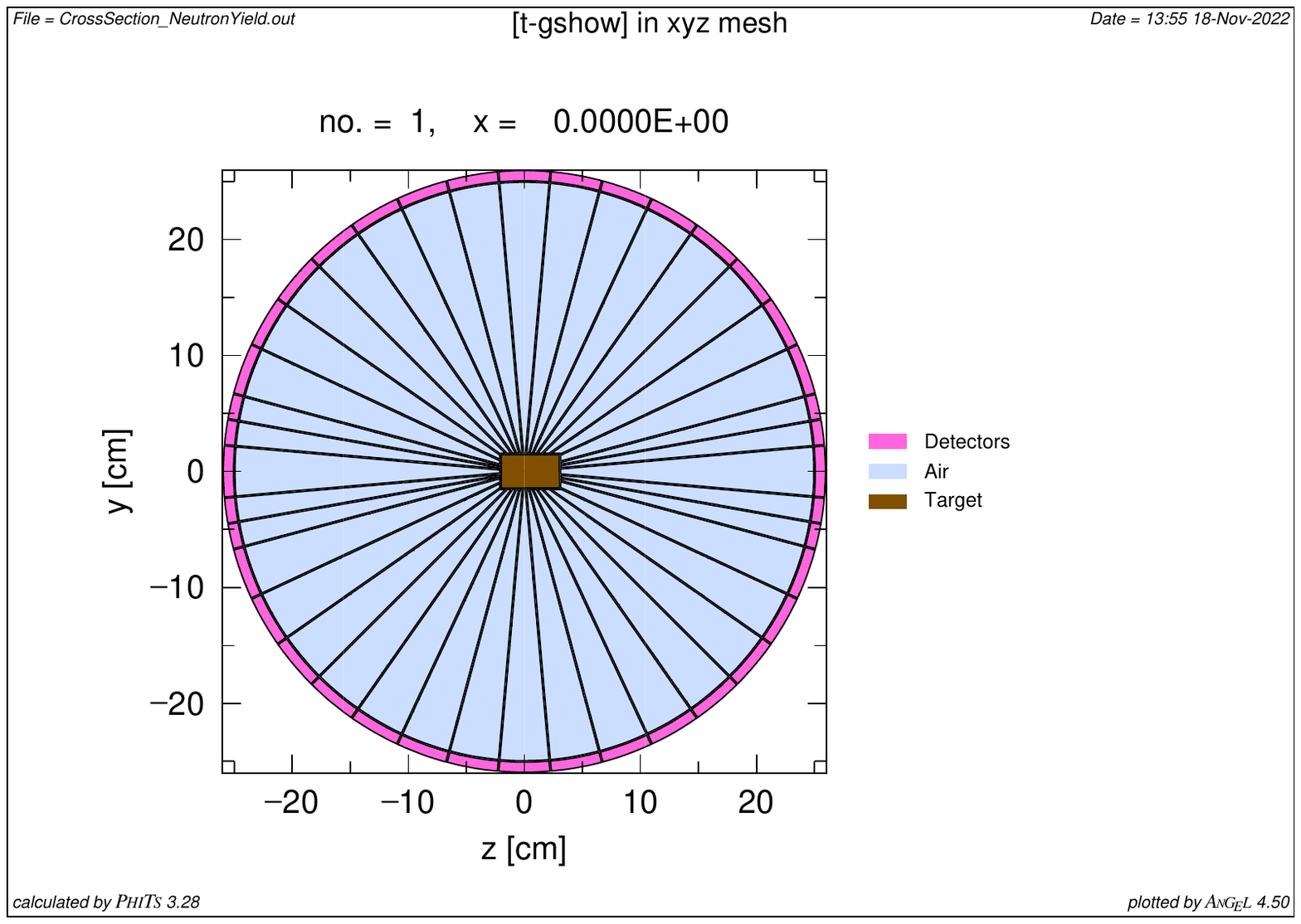}}
        \caption[Monte Carlo simulation geometry.]{Simulation description plots. They are created using PHITS Angel tool.}
    \end{figure}

	The observable is the double differential neutron yield $\mathfrak{F}$ which is determined by measuring the number of neutrons crossing a detector surface.
    We implemented several detectors as rings around the converter to resolve the scattering angle.
	This is illustrated in \autoref{fig:MC-3d-setup} and \autoref{fig:MC-crossSection}, where the black lines indicate the boundaries for each detector. 
    We give the concrete parameters for the scattering angle (center of the detector) $\Theta$ and its size $\Delta\Theta$ in the \autoref{sec:detector-data} in table \ref{tab:detector-data}.
 
	The cylindrical particle beam is inserted at position $(0,0,-9)$ into positive $z$ direction with a radius of $r_0 = \SI{0.5}{\centi\meter}$.
	Energy and particle type of the incoming particles are subject to variation as indicated by \autoref{tab:paramVariations}.
	The target is implemented as a cylinder along the $z$-axis centered around $(0,0,0)$ with a radius $r=\SI{1.5}{\centi\meter}$ and a variable length as well as a variable material, as indicated by \autoref{tab:paramVariations}, too.
	\begin{table}
		\centering
		\caption{Monte Carlo input parameters. The maximum number of sampled parameters is given in the last column. Rectangular brackets indicate numerical intervals with the lower value left and the upper value right inside the bracket.}
		\begin{tabular}{lcr}\toprule
			Quantity   & Values & Steps \\ \midrule
			Projectile & Deuterons, Protons   & 2\\
			$E_\text{kin}$ / MeV & $\left[ 3, 97\right]$ & 56\\
			Element       & Li, LiF, Be, Va, Ta & 5\\
			Length / mm   & $\left[ 0.02, 104.45\right]$ & 356\\
			Angle / \si{\degree}   & $\left[0, 180\right]$  & 21 \\\bottomrule
		\end{tabular}
		\label{tab:paramVariations}
	\end{table}
    We ensured that target was long enough to stop the full bunch inside the target.
    It is possible, however, that the target is longer than necessary to include low neutron absorption and moderation. 
    This is important since non-mono energetic beams are to be investigated; this would otherwise yield an overestimation of the neutron yield at higher energies.
    The target cylinder's radius was fixed at \SI{2.5}{\centi\meter}. 
	We acquired a total of \num{54768} neutron yield simulations in the mentioned setup.
    \subsubsection*{Data preparation}
    We organized the data as follows: 
    We labeled the categorical data features according to a binary representation of the values, while the continuous data was scaled from 0 to 1. 
    The two different projectiles can then be labeled by a single bit:
	Deuterium is labeled with 1, and Hydrogen is labeled with 0.
    For the converter materials under investigation, this makes it necessary to have 3 bits to describe all possibilities fully:
	'Li'--'000',  'LiF'--'001', 'Be'-- '010', 'Va'--'011', 'Ta'-- '100'

    Without preparation, the model training does not converge.
    The data was first resampled to increase the density of the data points in the higher energy part of the spectrum, which was logarithmically sampled from the code.
    This resampling was done with cubic spline interpolation of the Monte Carlos output data and joined with the raw data, doubling the number of data points for learning while keeping the spectral shape intact.
    
    Numerical stability is an issue because of the extensive range of input data, varying by several orders of magnitude. 
    While some data points contribute much to the mean squared error training quantity, some with a lower numerical value did not contribute much. 
    As a result, the training process was volatile, which needed to be mitigated.
    We fixed this by rescaling the data using the relations given in \autoref{eq:rescaling}.         
    \begin{equation}
        \begin{aligned}
            \Bar{Y} &= \log_{10}\left(Y\right)\\ 
            \widehat{Y} &= \Bar{Y} / \min\left( \Bar{Y} \right) \label{eq:rescaling}
        \end{aligned}
    \end{equation}
    where $\hat{Y}$ is the renormalized and $Y$ the raw data.
    Not at this point, the output data of the simulation is normalized to the incident particle, with the condition that the energy is below the spallation threshold, all values are smaller than 1, such that the value of the logarithm is negative.
    The final data reached from 1 to the normalized maximal value, and all data was larger than 0.
    The neutron's energy was also normalized by the cut-off energy to scale from 0 to 1.
    
    \subsection{ANN setup}
    We trained two models, one for the spectrum and one for the cut-off energy. 
    Training cut-off positions together with the spectrum are difficult since a cut-off represents a discontinuity in the spectrum.
    Discontinuities are not smooth and can only be approximated with an even more extensive data set. 
    The spectral shape is then normalized from minimal data at 0 to maximal non-zero data at 1, while the cut-off model predicts the relevant scaling for the predicted spectrum.

    The networks are both fully connected feed-forward networks.
    For the spectrum, the node structure is as follows, where $x_s$ is an 8-dimensional input vector:
    $(x_s\rightarrow 200 \rightarrow 200 \rightarrow 200 \rightarrow 200 \rightarrow 200 \rightarrow 200 \rightarrow 200 \rightarrow 1)$.
    For the cutoff energy of the spectrum, the node structure is as follows, where $x_c$ is a 7-dimensional input vector:
    $(x_c\rightarrow 260 \rightarrow 180 \rightarrow 180 \rightarrow 340 \rightarrow 180 \rightarrow 180 \rightarrow 340 \rightarrow 340  \rightarrow 180  \rightarrow 1)$.
    Since the spectrum model yields a full spectrum, an additional parameter marking the position inside the spectrum must also be used. 
    
    Better results might be acquired by dropout optimization or other sophisticated methods, but time constraints did not permit further optimizing the topology.
    The selected hyperparameters are given in \ref{sec:ANNhyperparams}
    
    \subsection{ANN bootstrap\label{sec:uncertainties}}
    The simulation data has, due to its probabilistic nature, an uncertainty. 
    This uncertainty is up to \SI{100}{\percent} for some of the energy bins and cannot be taken into account directly.
    We circumvented this with a bootstrap method\cite{Diaconis1983,Efron1981}. 
    The raw data was randomly resampled in the range of uncertainty. 
    We assumed a Gaussian distribution because the code also assumed it, yielding the uncertainty in a $1\sigma$ interval. 
    Then with this distribution function, we can resample the yield data $N$ times:
    \begin{equation}
        Y_\text{res} = \mathbf{G}\left( \mu = Y, \sigma = \sigma_Y\right),
    \end{equation}
    where $\mathbf{G}$ is the Normal distribution.
    Applying the same procedure on each resampled spectrum yields $N+1$ different predictions, which deviate around the joined mean.
    For each set of input parameters, $N+1$ predictions are generated. 
    The mean of these predictions is then the predicted value, while the standard deviation of those predictions is then the uncertainty of the prediction.

    \subsection{Ensuring generality}
    The formal application of the resulting surrogate $\mathfrak{F}$ is given by
    \begin{align}
    	\frac{\operatorname{d}^2\!Y}{\operatorname{d}\!\Theta\operatorname{d}\!E_n } &= \sum_{E_\text{p}}N_\text{p}\left(E_\text{p}\right)\cdot \hat{\mathfrak{F}}\left(E_n, \Theta, E_\text{p} \right)  \Delta E_\text{p}
     \intertext{and with the infinitesimal limit for $\Delta E_p \to 0$:}
    	&= \int N_\text{p}\left(E_\text{p}\right)\cdot \hat{\mathfrak{F}}\left(E_n, \Theta, E_\text{p} \right) \operatorname{d}\!E_\text{p},
    \end{align}
    where $Y$ is the neutron yield, $\Theta$ the scattering angle, $E_n$ the energy of the neutron, $N_\text{p}$ the number of incoming projectiles, $E_\text{p}$ the energy of the projectile, and $\hat{\mathfrak{F}}$ is the unit surrogate model, which is normalized to one projectile particle. 
    $E_\text{p}$ is sampled according to the needed precision. 
    We used a bin width of $\Delta E_\text{p} = \SI{1}{\mega\electronvolt}$.
    Setting the model up like this enables us to apply and verify this model on mono-energetic spectra and laser-accelerated ions.
    This model ensures direct comparability between conventional accelerator-driven ion sources and laser-driven ion sources.
    
\section{Results}
    The training of all networks converged when the learning rate was consequently reduced. 
    For each network, we extracted the mean squared error value and the normalization factor used during preprocessing and displayed them in \autoref{tab:trainingmetrics}
    \begin{table}[]
        \caption{Network metrics for all trained networks. Id 0 means the raw data is used. The other nine numbers indicate the nine resampled datasets. MSE is the mean squared error training metric, and Normalization indicates the normalization constant applied to each data set after the logarithm was applied.}
        \centering
        \begin{tabular}{ccc}\toprule
            Id & MSE & Normalization \\ \midrule
            0 & 0.00128 & -16.0505 \\
            1 & 0.00124 & -16.2796 \\
            2 & 0.00128 & -15.9784 \\
            3 & 0.00119 & -16.6231 \\
            4 & 0.00114 & -16.9578 \\
            5 & 0.00114 & -16.8753 \\
            6 & 0.00125 & -16.1653 \\
            7 & 0.00120 & -16.5274 \\
            8 & 0.00140 & -15.3819 \\
            9 & 0.00121 & -16.4534 \\ \bottomrule
        \end{tabular}
        \label{tab:trainingmetrics}
    \end{table}
    
    \subsection{Validation against the Raw Simulation Data\label{sec:valid-Sim}}
\begin{figure}
		\centering
		\subfloat[(p,Li) reaction under \SI{0}{\degree}.\label{fig:spectrumResult1a}]{\includegraphics[width=0.4\textwidth]{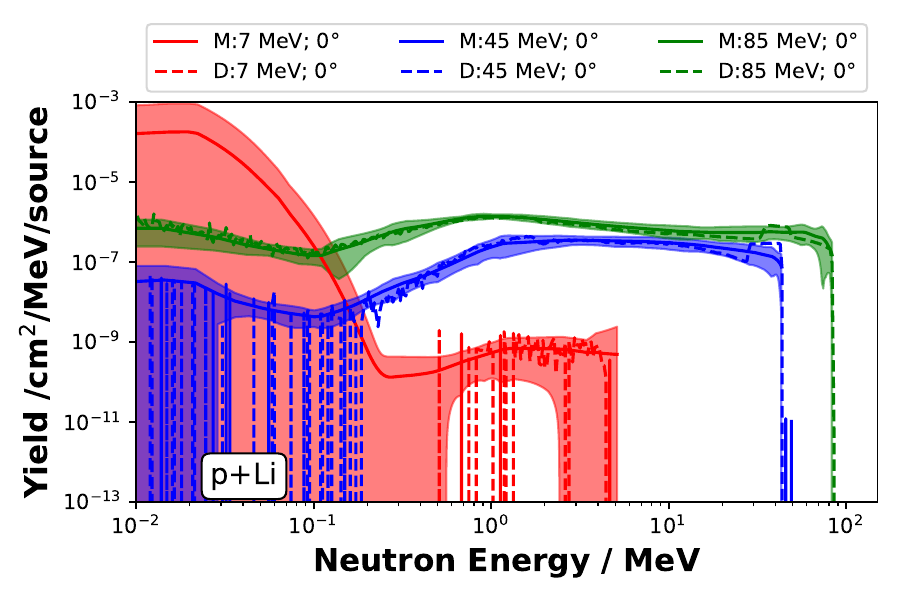}}\\
		\subfloat[(p,Li) reaction under \SI{90}{\degree}.]{\includegraphics[width=0.4\textwidth]{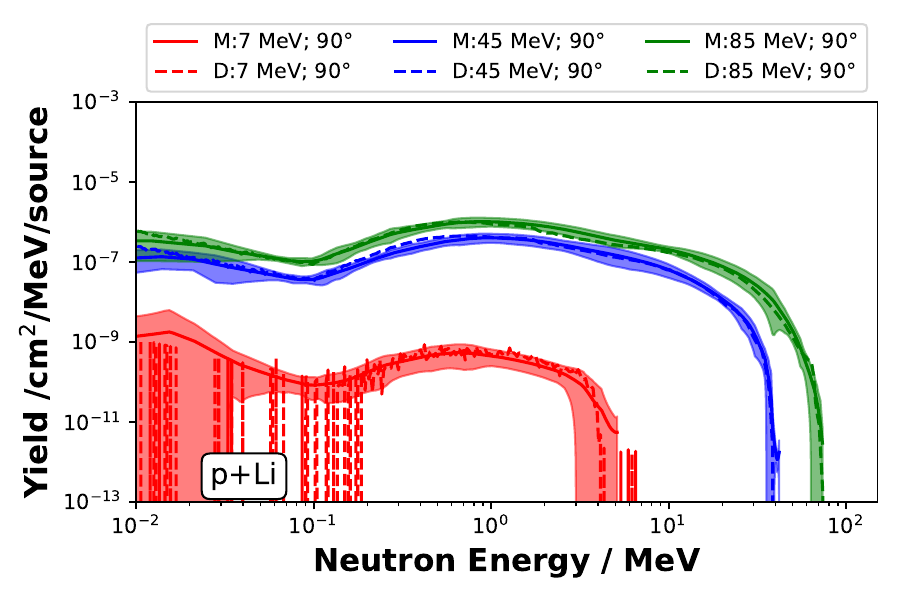}}\\
        \caption[Monte Carlo data vs. Surrogate prediction for proton projectiles]{Comparison of our model with results from PHITS. Displayed are samples for proton-induced neutron production on lithium. Dashed lines indicate the Monte Carlo data indicated by D, and solid lines indicate the corresponding model output by M. The shaded area is the model's uncertainty.}
		\label{fig:spectrumResult1}
	\end{figure}

\begin{figure}
		\centering
		\subfloat[(p,Be) reaction under \SI{0}{\degree}.]{\includegraphics[width=0.4\textwidth]{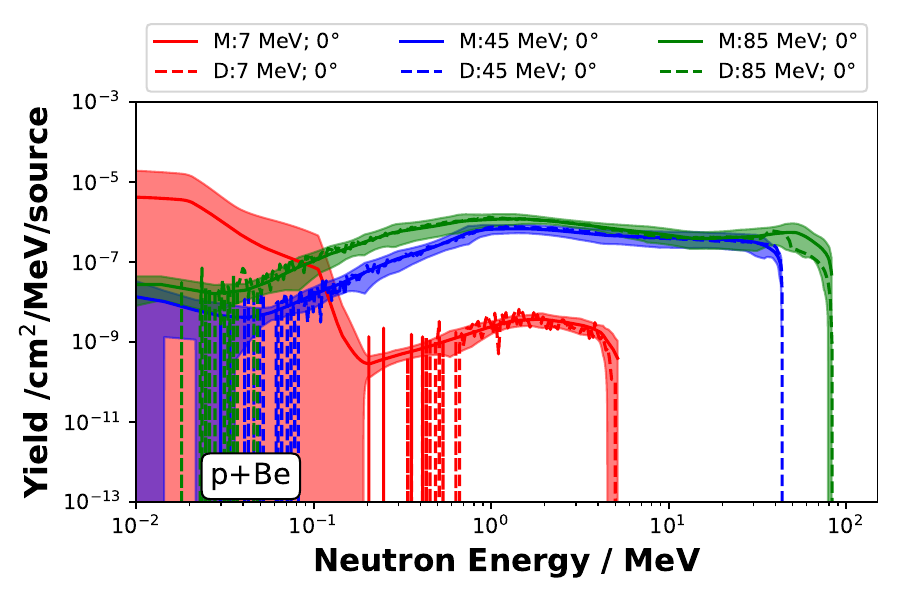}}\\
		\subfloat[(p,Be) reaction under \SI{90}{\degree}.]{\includegraphics[width=0.4\textwidth]{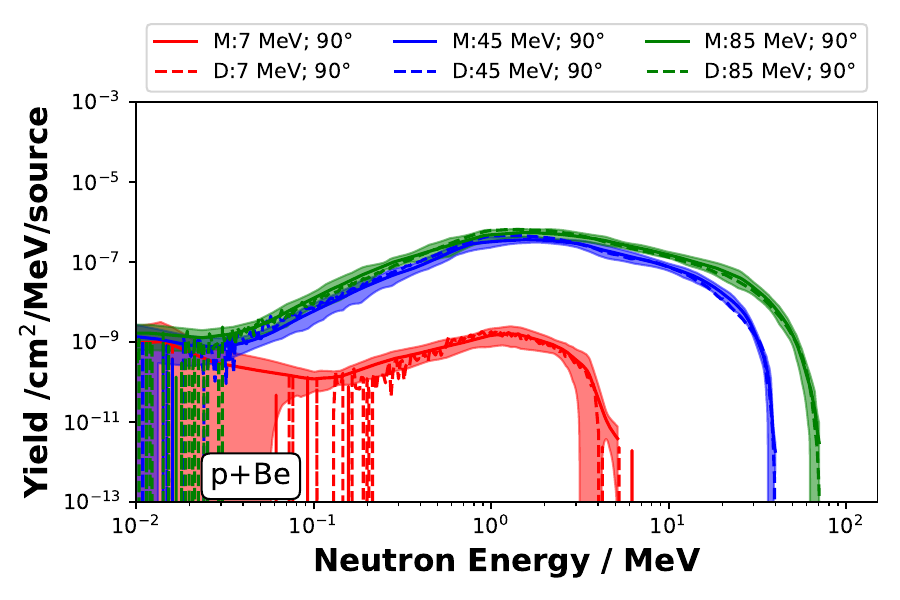}}\\
		\caption[Monte Carlo data vs. Surrogate prediction for proton projectiles]{Comparison of our model with results from PHITS. Displayed are samples for proton-induced neutron production on beryllium. Dashed lines indicate the Monte Carlo data indicated by D, and solid lines indicate the corresponding model output by M. The shaded area is the model's uncertainty.}
		\label{fig:spectrumResult2}
	\end{figure}

\begin{figure}
		\centering
		\subfloat[(d,Li) reaction under \SI{0}{\degree}.]{\includegraphics[width=0.4\textwidth]{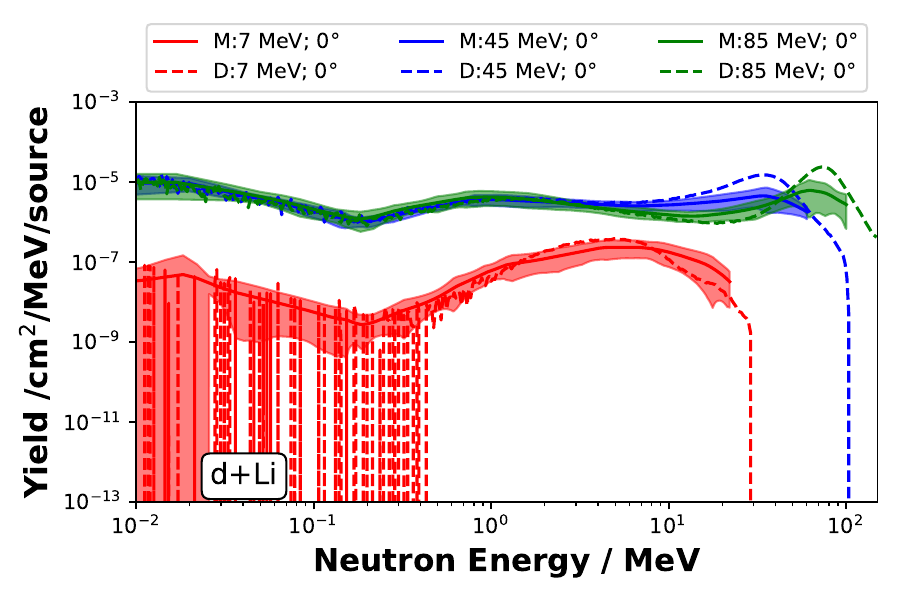}}\\
		\subfloat[(d,Li) reaction under \SI{90}{\degree}.]{\includegraphics[width=0.4\textwidth]{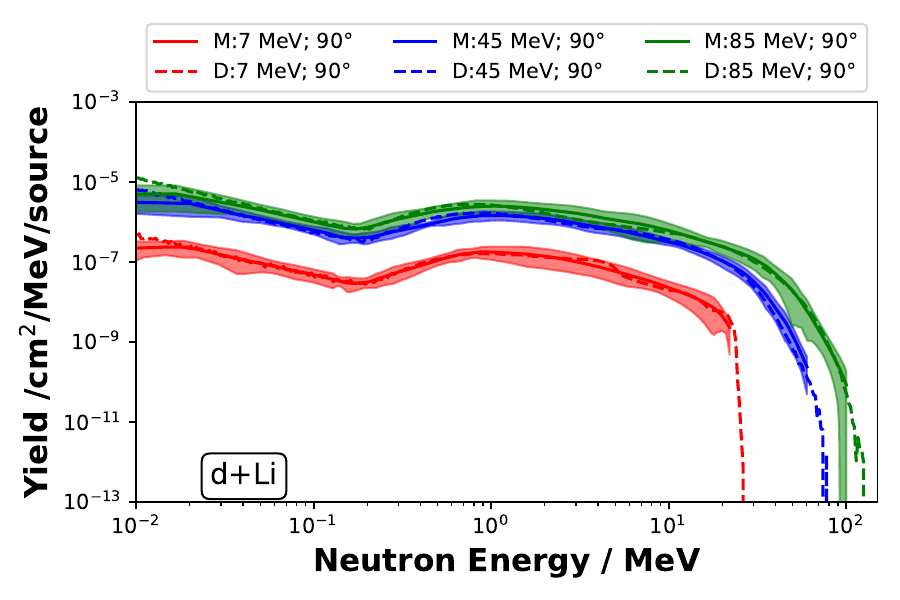}}
        \caption[Monte Carlo data vs. Surrogate prediction for deuterium projectiles.]{Comparison of our model with results from PHITS. Displayed are samples for deuterium-induced neutron production on lithium. Dashed lines indicate the Monte Carlo data indicated by D, and solid lines indicate the corresponding model output by M. The shaded area is the model's uncertainty.}
		\label{fig:spectrumResult3}
	\end{figure}

\begin{figure}
		\centering
		\subfloat[(d,Be) reaction under \SI{0}{\degree}.]{\includegraphics[width=0.4\textwidth]{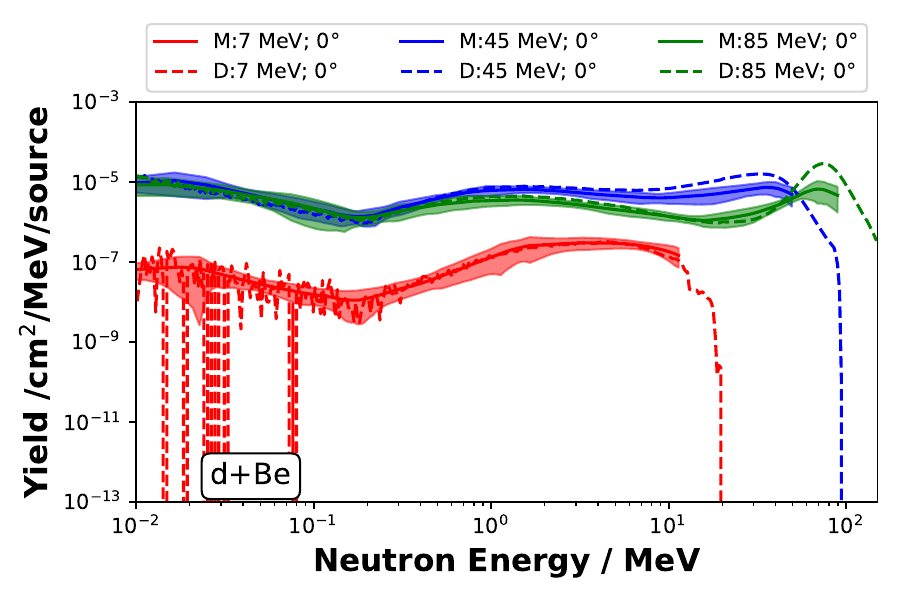}}\\
		\subfloat[(d,Be) reaction under \SI{90}{\degree}.]{\includegraphics[width=0.4\textwidth]{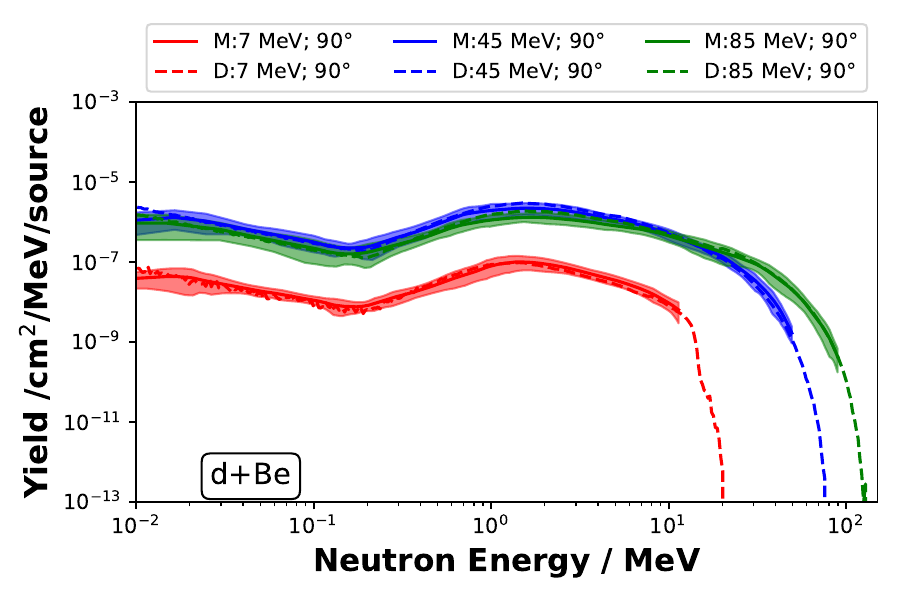}}\\
		\caption[Monte Carlo data vs. Surrogate prediction for deuterium projectiles.]{Comparison of our model with results from PHITS. Displayed are samples for deuterium-induced neutron production on beryllium. Dashed lines indicate the Monte Carlo data indicated by D, and solid lines indicate the corresponding model output by M. The shaded area is the model's uncertainty.}
		\label{fig:spectrumResult4}
	\end{figure}

    To check the validity of the surrogate, we first compare its predictions with the raw data used for training.
    The results are displayed in \autoref{fig:spectrumResult1},\autoref{fig:spectrumResult2}, \autoref{fig:spectrumResult3}, and \autoref{fig:spectrumResult4}.
    Displayed are the full spectra created by the Monte Carlo code vs the result created by our model, with its corresponding uncertainty.
    We can derive a few remarks on the model's validity from these figures.
	\begin{compactenum}
		\item The model is only valid in the region of fit, defined by the parameters in \autoref{tab:paramVariations}.
		\item  After the cutoff energy, the spectrum generally deviates, implying that the spectral model does not learn the physical cutoff and therefore yields wrong results in the high-energy end of the spectrum. 
		\item The model can predict the lowest energy neutrons, which can not be seen in the Monte Carlo spectra. These values, however, have to be used with care: Their uncertainty is very high, up to \SI{100}{\percent}, and the result does not reproduce measurement data correctly (confer to \autoref{sec:valid-Exp}).
		This has two reasons: First, the model itself is regression-based, and as can be seen, higher energy projectile energies do have neutron signals in the low energy regime. 
		The effect of the regression is calculating a corresponding mean value from all calculated spectra, yielding a non-zero contribution.
		Second, the reason for zero count rates is the bulky target, the lowest neutrons created by low-energy projectiles can not leave the target and are stuck there. 
        The count rates in the Simulation are low, meaning the statistical uncertainties are high.
		\item The proton lithium reaction has a characteristic high energy peak in the forward direction.
	    This peak is suppressed by the bulky target, as seen in \autoref{fig:spectrumResult1a}.
		Due to the nature of the regression model, the model further suppresses the peak and is, therefore, impossible to resolve.
        \item The cutoff energies mean squared error is approximately \SI{9}{\mega\electronvolt^2}. 
        This implies an uncertainty of the cut-off energy of approximately \SI{\pm 3}{\mega\electronvolt}.
	\end{compactenum}
	Despite these limitations, the model can predict accurate spectra over a large energy domain in milliseconds evaluation time and allows quick evaluations of neutron production.

    \subsection{Validation against Experimental Data\label{sec:valid-Exp}}
    The largest experimental data collection is the EXFOR database, hosted by the International Atomic Energy Agency (IAEA)\cite{OTUKA2014272,ZERKIN201831}.
    Only a few data sets are comparable to the model derived here due to different experimental setups, the focus on cross-section measurements with thin targets, and energies of interest larger than \SI{100}{\mega\electronvolt}.
    \begin{figure}
        \centering
        \subfloat[\label{fig:exp:kamada}]{\includegraphics[width=.48\textwidth]{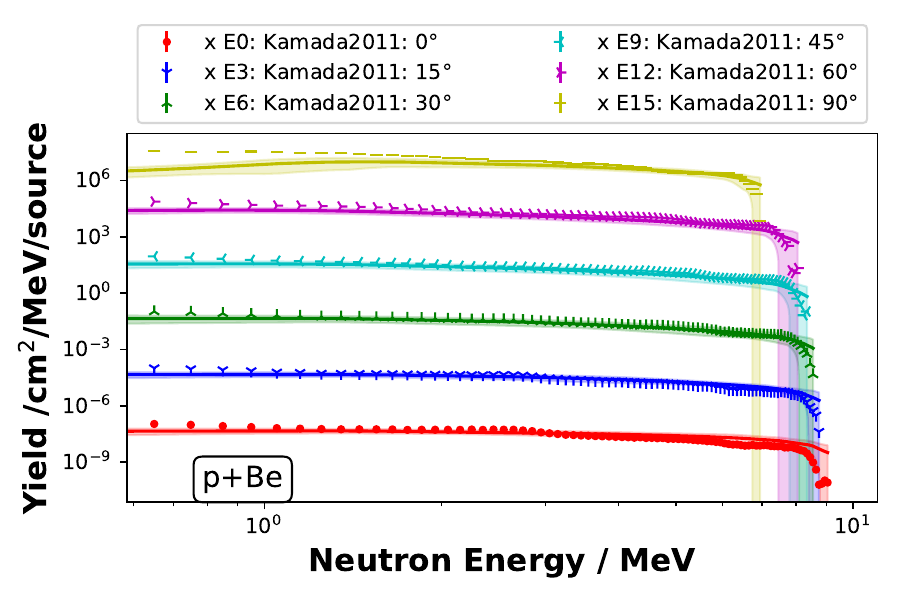}}\\
        \subfloat[\label{fig:exp:osipenko1}]{\includegraphics[width=.48\textwidth]{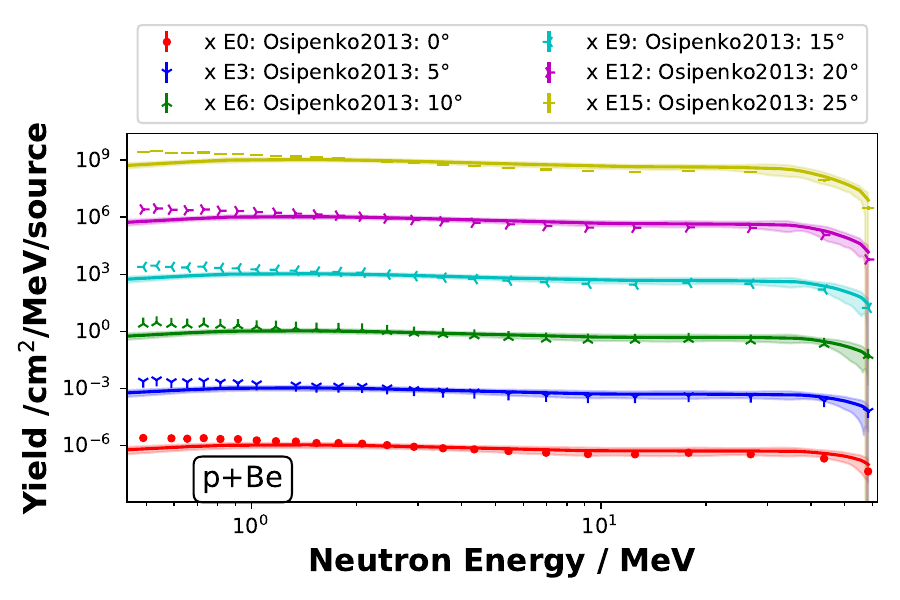}}\\
        \subfloat[\label{fig:exp:osipenko2}]{\includegraphics[width=.48\textwidth]{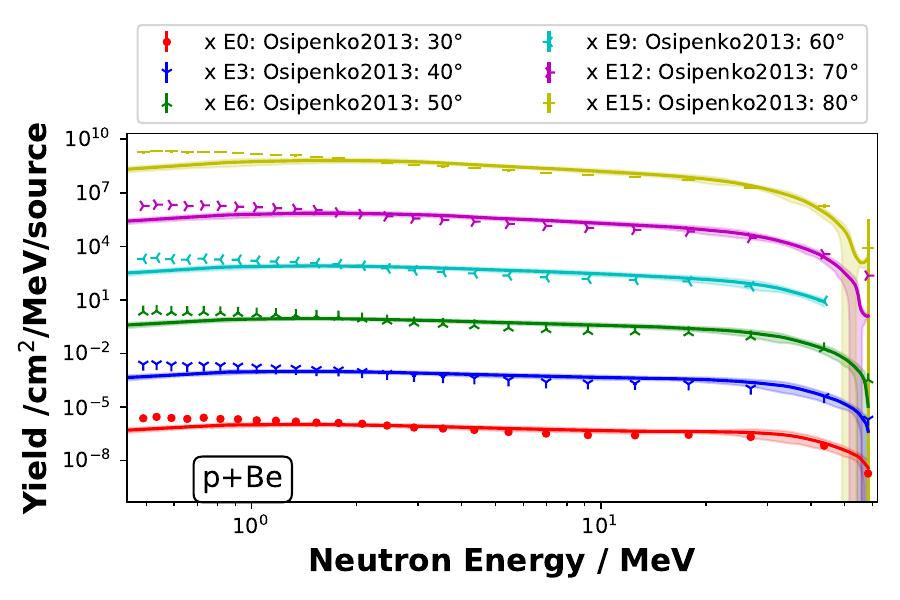}}
        \caption{Experimental data compared to the model (solid lines with shaded uncertainty). (a) Data taken by Kamada et al. in 2011 with $E_p = \SI{11}{\mega\electronvolt}$ \cite{Kamada2011}. (b) and (c) contain data taken by Osipenko et al. in 2013 with $E_p = \SI{62}{\mega\electronvolt}$ \cite{Osipenko2013}. 
        The scattering angle is listed in the legend, the model prediction is the same color as the data, and the data has been multiplied by a factor as indicated by the legends prefix to increase the readability of the plot.}
        \label{fig:ExpDataCompare}
    \end{figure}
    
    It is visible that the model, displayed in \autoref{fig:ExpDataCompare} deviates from the data in the low energy region below \SI{1}{\mega\electronvolt}. 
    The dataset from Howard et al. taken in 2001 with $E_p = \SI{3}{\mega\electronvolt}$\cite{Howard2001} could not be adequately reproduced and is therefore not displayed here. 
    Both observations have the same reason, for low energy neutrons, the count rates in the Monte Carlo code were too low; as such, the relative uncertainty is as high as \SI{100}{\percent}.

    \subsection{Spectra for different conventional setups}
    We can use the model to compare different setups to each other. 
    The variations do differ in the number of particles and the input energies. 
    Expectations are that higher ion energy correlates to a higher neutron count and higher cutoff energy of the spectrum.
    The number of neutrons is linearly dependent on the incoming particles due to the superposition principle.

\begin{figure}
        \centering
        \subfloat[Signal under \SI{0}{\degree}.\label{fig:conventional00}]{\includegraphics[width=.48\textwidth]{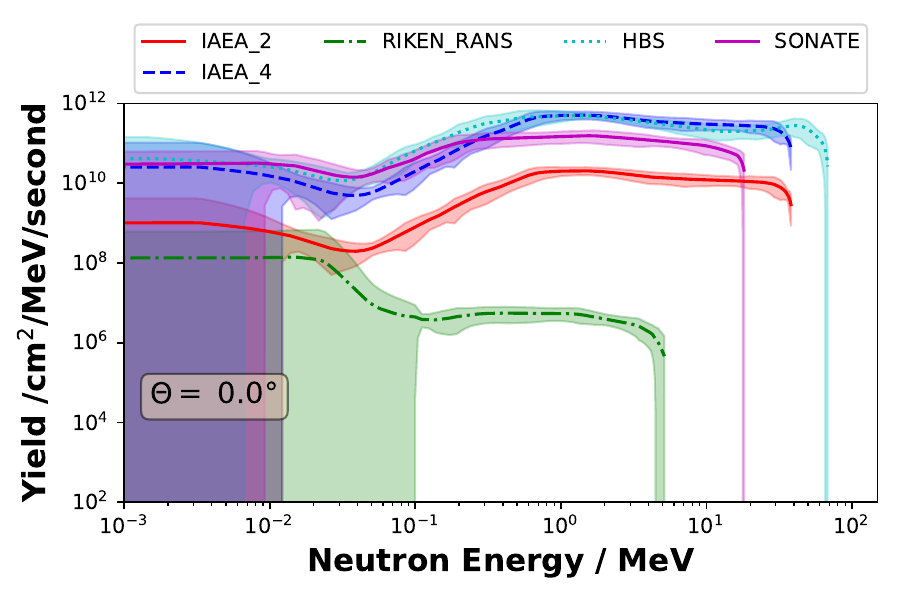}} \hfill
        \subfloat[Signal under \SI{90}{\degree}.\label{fig:conventional90}]{\includegraphics[width=.48\textwidth]{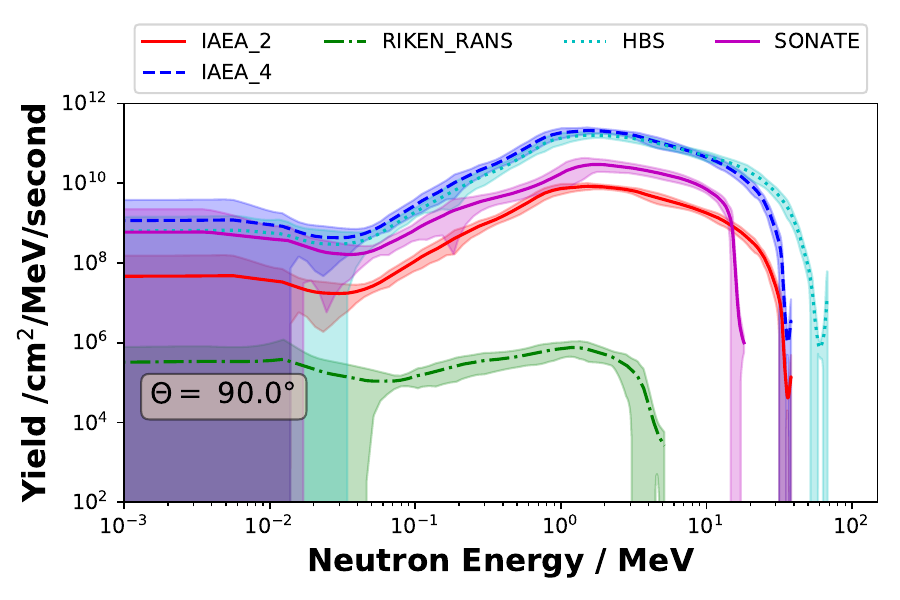}}
        \caption{Model output for the conventional accelerators in p+Be configuration.}
        \label{fig:conventionalCompare}
    \end{figure}

    The parameters for the conventional accelerators are extracted from several reports focused on developing compact ion-based neutron sources. 
    Namely the projects HBS in Jülich \cite{Ruecker2016}, RANS at RIKEN in Japan \cite{OTAKE2018}, SONATE at CEA-Saclay \cite{refId0}, and from the technical report of the IAEA in Vienna \cite{international2021iaea}.
    The parameters used for this evaluation are tabulated in \autoref{tab:CANSparametersPBe}, and the model's results are displayed in \autoref{fig:conventional00} and \autoref{fig:conventional90}.
	\begin{table}
		\centering
        \caption[Parameters of conventional neutron source projects.]{Parameters for several compact neutron source accelerators. 
        The same projectile and target types are fixed as protons and beryllium to compare different spectra. The sources are given in the text.}
		\begin{tabular}{crrr}\toprule
			Name &  Energy & Current  & $d_T$  \\
            Unit & MeV & \SI{E-4}{\ampere} &  mm\\\midrule
            IAEA2 & 40 & 50 & 20 \\
            IAEA4 & 40 & 1250 & 20 \\ 
            RANS  &  7 &  1 & 0.3 \\ 
            HBS   & 70 &  1000 & 16 \\ 
            SONATE & 20 & 1000 & 2 \\ \bottomrule
		\end{tabular}
		\label{tab:CANSparametersPBe}
	\end{table}
    Differences in the spectrum result from the different particle energies and level differences from the different particle counts. 
    The spectral shape also depends on the scattering angle.

    \subsection{Spectra for Laser Accelerated Protons}
\begin{figure}
        \centering
        \subfloat[Signal under \SI{0}{\degree}.\label{fig:TNSA00}]{\includegraphics[width=.48\textwidth]{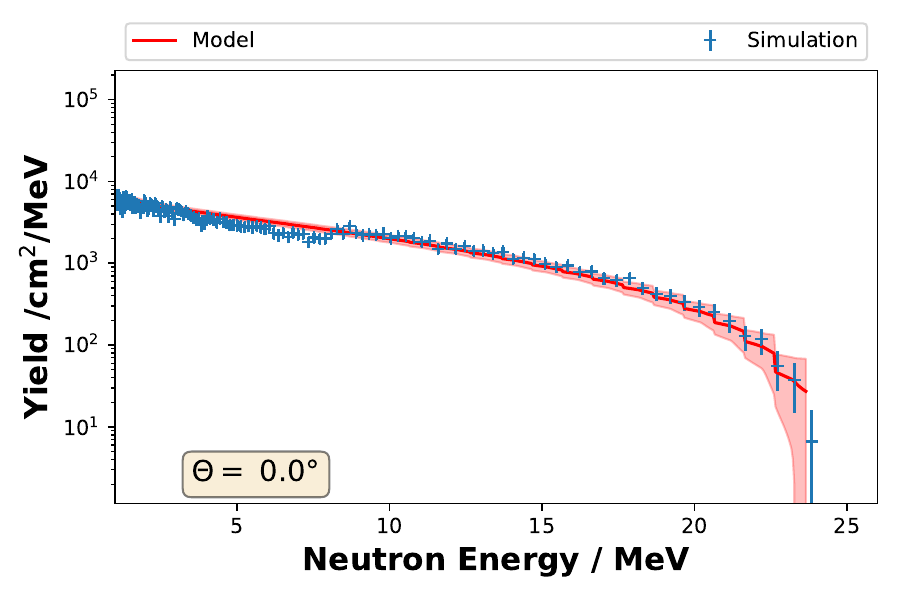}} \hfill
        \subfloat[Signal under \SI{60}{\degree}.\label{fig:TNSA60}]{\includegraphics[width=.48\textwidth]{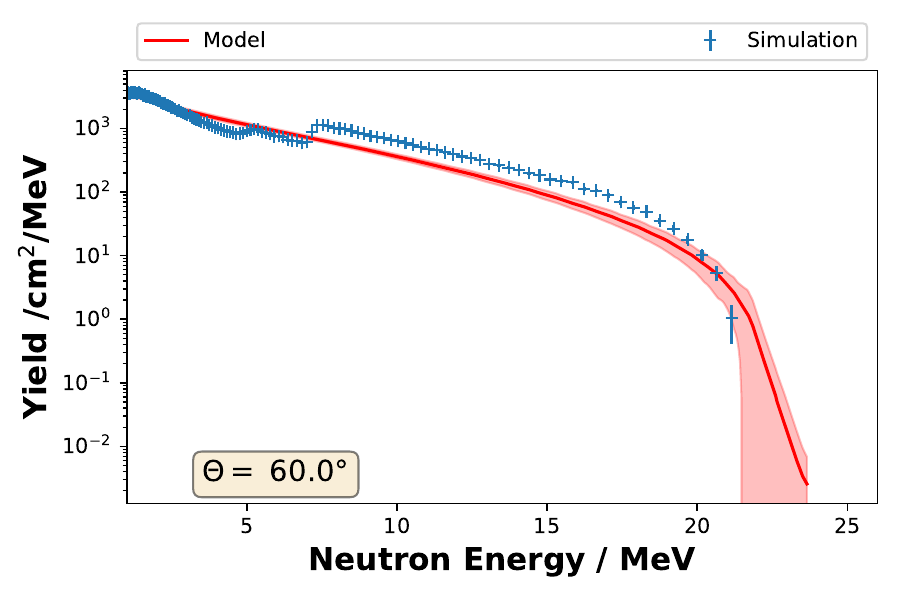}}
        \caption{Model and simulation output for a TNSA proton beam.}
        \label{fig:TNSASpectrum}
    \end{figure}

    As mentioned previously, the laser-accelerated spectra are calculated by the superposition of several mono-energetic models which follow the composition of the full laser particle spectrum.
    We tested this on a simulation for the ion spectrum evaluated by Schmitz et al.\cite{Schmitz2022}.
    The spectrum is analytically given as 
    \begin{equation}
        \frac{\operatorname{d}\!N}{\operatorname{d}\!E}(E) = \frac{N_0}{x} \exp\left( - \frac{x}{k_B T} \right),
    \end{equation}
    where $N_0 = \num{3.7E+11}$, $k_B T = \SI{7.4}{\mega\electronvolt}$, and $E$ the spectrum's energy in the interval from \SI{5}{\mega\electronvolt} to \SI{25.5}{\mega\electronvolt}.
    We did additional simulations for this proton spectrum and a corresponding beryllium target with a thickness of \SI{5.02}{\milli\meter}.
    The results of the simulations, together with the corresponding model output, are displayed in \autoref{fig:TNSASpectrum}.
    The model output is close to the simulation at \SI{0}{\degree}. 
    Further deviation towards higher scattering angles causes a larger model deviation from the simulation, indicated by a step in the spectrum at around \SI{7}{\mega\electronvolt}.
    This step increases with a larger scattering angle, which indicates a dependency on the underlying geometrical structure of the converter.
    With a larger angle, the neutron's range inside the converter increases since the transverse dimension of the target is much larger than the thickness of the target in beam direction (\SI{240}{\milli\meter} vs. \SI{5.02}{\milli\meter}). 
    This model's smooth behavior is caused by its nature of being a regression model. 
    It smoothes out the spectrum while taking deviations from all angles into account.
    
\section{Conclusion}
    We presented a predictive surrogate model using an extensive Monte Carlo simulation data set.
    This model can be deployed quickly and provide the core for quick evaluations of neutron yields, increasing the accessibility of neutron yield simulations and minimizing the need for experience and training for the complex Monte Carlo tools.
    Despite insufficient experimental data for data-driven modeling, our model achieves a good agreement between the model and experimental datasets.

    Due to our simulation setup and its numerical feasibility, neutrons from projectile energies $E_p < \SI{5}{\mega\electronvolt}$ have a large uncertainty if $E_n < \SI{1}{\mega\electronvolt}$.
    We recommend not trusting the model in this region.
    The uncertainty, which we calculate by the already explained bootstrapping method in \autoref{sec:uncertainties} is a good indicator of the model's validity.

    The next step to improve this model is to investigate the geometry dependency closer and to increase the cut-off modeling for the neutron spectra.
    Both could be achieved by new experimental data or an extension of the simulation model. 

\section*{Author Declarations}
The authors have no conflicts of interest to disclose. 
\section*{Data availability statement}
Codes\footnote{\url{https://git.rwth-aachen.de/surrogat-models/surrogate-modeling-for-neutron-production}} and data\footnote{\url{https://git.rwth-aachen.de/surrogat-models/neutron-data}} are available in public repositories.
\section*{Funding Statement}
This work was funded by HMWK through the LOEWE center “Nuclear Photonics.”

This work is also supported by the Graduate School CE within the Centre for Computational Engineering at Technische Universität Darmstadt.

\section*{Acknowledgments}
We thank Moritz von Treskow and Dimitrios Loukrezis for the helpful discussions on the data pipeline setup and their thoughts on uncertainty quantification for Artificial Neural Networks.

\appendix
\section{Detector Details \label{sec:detector-data}}
    The radius of the detector sphere is \SI{25}{\centi\meter}. 
    The area of the detector is the absolute covered area of the specific detector, and the Ratio is the percentage of full coverage. 
    $\Theta_\text{min}$ and $\Theta_\text{max}$ values describe the minimum and maximum scattering angle the detector can measure.
    $\Theta_\text{mean}$ is the central scattering angle of the detector.
\begin{table}
        \centering
        \caption{Data for detectors implemented in the Monte Carlo model. Every second value is grayed to guide the eye.}
        \begin{tabular}{crrrrrr}
            \toprule
              & $\Theta_\text{min}$ & $\Theta_\text{max}$ & $\Theta_\text{mean}$ & $\sigma\Theta$ & Area & Ratio \\ 
              & $\si{\degree}$ & $\si{\degree}$ & $\si{\degree}$ & $\si{\degree}$ & $\si{\centi\meter\squared}$& \si{\percent}  \\ \midrule
                  D01 &   0&   5&   0.0 & 5.0 &  14.94 & 0.19\\
\rowcolor{gray!20}D02 &   5&  10&   7.5 & 2.5 &  44.72 & 0.57\\
                  D03 &  10&  15&  12.5 & 2.5 &  74.15 & 0.94\\
\rowcolor{gray!20}D04 &  15&  25&  20.0 & 5.0 & 234.12 & 2.98\\
                  D05 &  25&  35&  30.0 & 5.0 & 342.26 & 4.36\\
\rowcolor{gray!20}D06 &  35&  45&  40.0 & 5.0 & 440.00 & 5.60\\
                  D07 &  45&  55&  50.0 & 5.0 & 524.37 & 6.68\\
\rowcolor{gray!20}D08 &  55&  65&  60.0 & 5.0 & 592.81 & 7.55\\
                  D09 &  65&  75&  70.0 & 5.0 & 643.24 & 8.19\\
\rowcolor{gray!20}D10 &  75&  85&  80.0 & 5.0 & 674.12 & 8.58\\
                  D11 &  85&  95&  90.0 & 5.0 & 684.52 & 8.72\\
\rowcolor{gray!20}D12 &  95& 105& 100.0 & 5.0 & 674.12 & 8.58\\
                  D13 & 105& 115& 110.0 & 5.0 & 643.24 & 8.19\\
\rowcolor{gray!20}D14 & 115& 125& 120.0 & 5.0 & 592.81 & 7.55\\ 
                  D15 & 125& 135& 130.0 & 5.0 & 524.37 & 6.68\\
\rowcolor{gray!20}D16 & 135& 145& 140.0 & 5.0 & 440.00 & 5.60\\
                  D17 & 145& 155& 150.0 & 5.0 & 342.26 & 4.36\\
\rowcolor{gray!20}D18 & 155& 165& 160.0 & 5.0 & 234.12 & 2.98\\
                  D19 & 165& 170& 167.5 & 2.5 &  74.15 & 0.94\\
\rowcolor{gray!20}D20 & 170& 175& 172.5 & 2.5 &  44.72 & 0.57\\
                  D21 & 175& 180& 180.0 & 5.0 &  14.94 & 0.19\\            \bottomrule
        \end{tabular}
        \label{tab:detector-data}
    \end{table}

\section{Hyperparameters\label{sec:ANNhyperparams}}
We used the Rectified Linear Unit (ReLU) for the activation function, which is widely used for Regression problems.
The identity is employed for the activation function for the last step to the output layer.
Similarly, we chose the mean squared error, suited for regression problems, as loss, and it was minimized using the Adam optimizer with $\beta_1 = 0.9$, $\beta_2 = 0.999$ and $\epsilon = \SI{1e-07}{}$. 
The initial learning rate was 0.001, which was lowered to a minimum of 0.0001 during training should the optimizer detect a plateau in the validation loss value (Keras' ReduceLROnPlateau feature). 

Considering the resampling, each simulation output contains information about $\sim 500$ locations (400 raw data + 100 resampled data) in the energy spectrum. 
Hence, the available data length for the continuous spectral model was $\num{54768} \times 400 = \num{21907200}$ data points. 
Of these, we used 81\% for training, 9\% 
Every training used a batch size of 16 and an early stopping mechanism. 
Each hidden layer in the network has an L1 regularization strength of \num{3.56e-07} and an L2 regularization strength of \num{1e-14}.
Since the maximum energy is only predicted per simulation and not per energy bin of the energy spectra, the model for the cut-off energy was trained on \num{54768} unique data points. 
This significantly smaller dataset made the model training faster.
The L1 regularization strength was set to \num{6.2613e-7}, and the L2 regularization strength was set to \num{2.0392e-8}.

\bibliographystyle{elsarticle-num} 
\bibliography{bibliography-neutron}

\end{document}